\documentclass[twocolumn]{aastex631}

\usepackage{amsmath}
\usepackage{mathtools}

\begin{document}


\title{Introduction of Machine Learning for Astronomy (Hands-on Workshop)}

\author{Yu~Wang}
\affiliation{ICRA, Dip. di Fisica, Universit\`a  di Roma ``La Sapienza'', Piazzale Aldo Moro 5, I-00185 Roma, Italy}
\affiliation{ICRANet, Piazza della Repubblica 10, I-65122 Pescara, Italy} 
\affiliation{INAF -- Osservatorio Astronomico d'Abruzzo,Via M. Maggini snc, I-64100, Teramo, Italy.}

\author{Rahim Moradi}
\affiliation{ICRA, Dip. di Fisica, Universit\`a  di Roma ``La Sapienza'', Piazzale Aldo Moro 5, I-00185 Roma, Italy.}
\affiliation{ICRANet, Piazza della Repubblica 10, I-65122 Pescara, Italy} 
\affiliation{INAF -- Osservatorio Astronomico d'Abruzzo,Via M. Maggini snc, I-64100, Teramo, Italy}

\author{Mohammad H. Zhoolideh Haghighi}
\affiliation{Department of Physics, K.N. Toosi University of Technology, Tehran P.O. Box 15875-4416, Iran}
\affiliation{School of Astronomy, Institute for Research in Fundamental Sciences (IPM), Tehran, 19395–5746, Iran}

\author{Fatemeh Rastegarnia}
\affiliation{ICRANet, Piazza della Repubblica 10, I-65122 Pescara, Italy}
\affiliation{Dip. di Fisica e Scienze della Terra, Universit\`a degli Studi di Ferrara, Via Saragat 1, I--44122 Ferrara, Italy}
\affiliation{Department of Physics,Faculty of Physics and Chemistry, Alzahra University,Tehran, Iran}

\email{yu.wang@icranet.org, rahim.moradi@icranet.org, \\mzhoolideh@ipm.ir, f.rastegarnia@alzahra.ac.ir}

\begin{abstract}
This article is based on the tutorial we gave at the hands-on workshop of the ICRANet-ISFAHAN Astronomy Meeting\footnote{\url{https://indico.icranet.org/event/2/page/12-data-science-in-relativistic-astrophysics-hands-on-workshop}}. We first introduce the basic theory of machine learning and sort out the whole process of training a neural network. We then demonstrate this process with an example of inferring redshifts from SDSS spectra. To emphasize that machine learning for astronomy is easy to get started, we demonstrate that the most basic CNN network can be used to obtain high accuracy, we also show that with simple modifications,  the network can be converted for classification problems and also to processing gravitational wave data.
\end{abstract}





\section{Introduction}\label{sec:introduction}

Machine learning is no more a new concept as currently it is involved in many aspects of life. Every time you pick up the phone and unlock it using face recognition and translate your voice into text when chatting, all have contributions from machine learning and in terms of the professional field, the first time that machine learning made a major contribution was in 2016, when Google's AlphaGo playing Go defeated the world champion LiShishi and Ke Jie \citep{2016Natur.529..484S, 2017Natur.550..354S, 2018Sci...362.1140S, 2020Natur.588..604S}, Alphago played some of the moves that confused human beings and these are commonly used now in
 human-human combat. If you often play go, you will find that many human players play with a machine style. In other words, humans accept the changes brought by machines. Training alphaGo used convolutional neural networks \citep{fukushima1982neocognitron, lecun1989backpropagation}. Today our examples use a similar network. Last year  Google's Deepmind team announced that protein folding prediction could also be successfully solved in this way and this had posed a major problem in biology for almost 50 years.  \citep{2021Natur.596..583J}. Also they immediately used the trained machine to predict the protein structure of the COVID19 virus to help in the development of drugs \citep{arora2020artificial,jumper2021applying}.

\begin{figure*}
\centering
\includegraphics[width=1\hsize,clip]{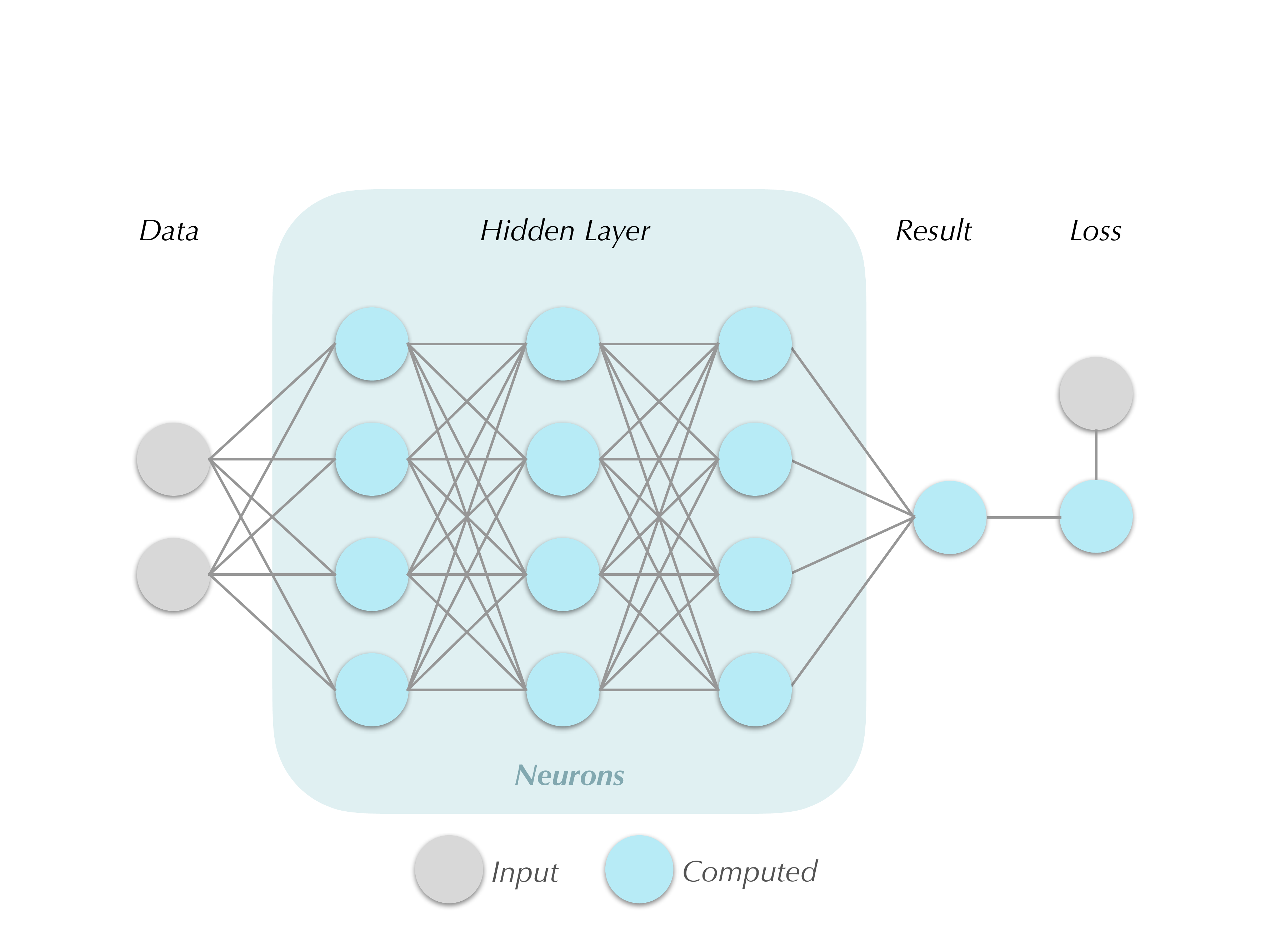}
\caption{Neural network composed of various neurons. features and labels are the input data represented by grey dots, the hidden layers constructed by neurons and activation functions are represented by blue dots. }
\label{fig:neuralnetwork}
\end{figure*}

Returning to astronomy, its data records spatial and temporal information, which is very close to image recognition and speech recognition. The pictures of galaxies seen by telescopes, the spectra of detected radiation, and the structure of the universe generated by simulations, all these can be categorized into image analysis. The evolution of celestial bodies, the lightcurves of explosions, and simulations of changes at each time step, can be used with techniques similar to voice recognition. Broadly speaking, the only axes of the four-dimensional universe we live in are time and space, so the data is no exception with time and space. What machine learning has to do is to receive this data and then give an answer.

Abstracted to mathematics, machine learning can be seen as a map, mapping data to answer. For deep learning \citep[see e.g.][and references therein]{lecun2015deep,goodfellow2016deep,dong2021survey}, this map consists of many layers of neurons. Deep learning is a subset of machine learning, which has developed rapidly in the last decade. Machine learning is also a subset of artificial intelligence (AI), which has been talked about for many years. The AI that surpasses human beings  in science fiction movies is defined as strong AI \citep{strauss2018big, butz2021towards}, the development of strong AI has only reached the stage of thesis. But according to our estimation, now the computer can probably simulate millions of neurons. The human brain has hundreds of millions of neurons. According to the speed of computer development, when in about 20 years, the computer can reach the complexity of the human brain, then strong AI will be really practical.

This article is based on our hands-on tutorial in the workshop. In section \ref{sec:basics} we introduce the principle and implementation of neural networks, in section \ref{sec:examples}  we illustrate it with an example of inferring redshift by SDSS spectra, and in section \ref{sec:extensions} we demonstrate that the application can be extended to classification and other sources by simple changes to the neural network. This article does not cover the code explanation, running and troubleshooting in the workshop.

\section{Basics}\label{sec:basics}

\begin{figure*}
\centering
\includegraphics[width=1\hsize,clip]{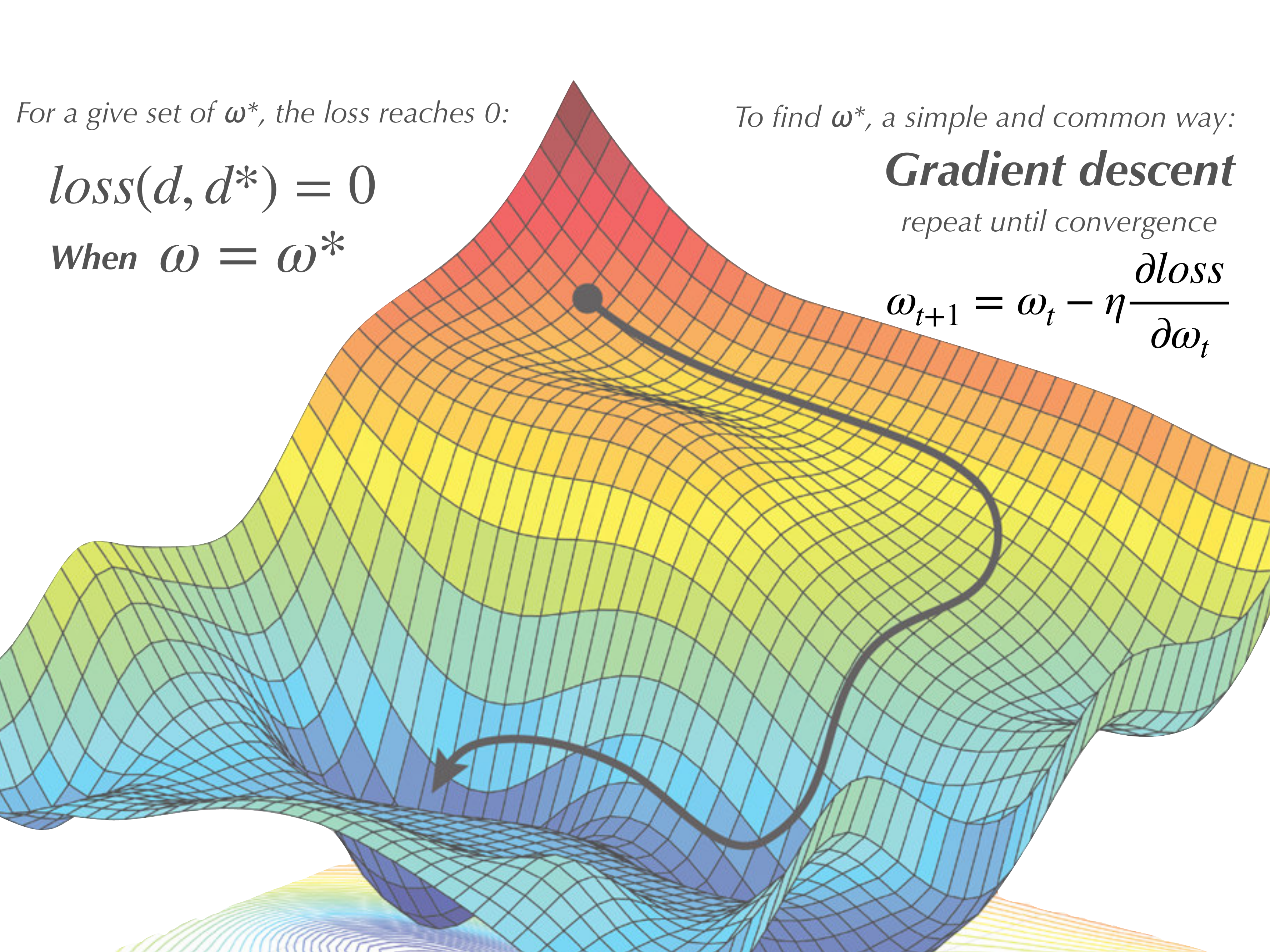}
\caption{Demonstration of the gradient descent process, the loss decreases following the curve line from the red region to the blue valley via the change of $\omega$ at each training step, eventually, the $\omega$ approaches $\omega^*$. This figure is reproduced from \citep{amini2018spatial}.}
\label{fig:gradientdescent}
\end{figure*}

Figure \ref{fig:neuralnetwork} shows an example of a neural network. The two leftmost grey dots represent the input, which is passed into the middle neural network, composed of three layers, and each layer contains four neurons. If we know the correct answer, we can define a loss function to compare the difference between the machine's answer and the real answer. For example, if we input the LIGO data, we can calculate whether it contains the gravitational wave signal, and the parameters of the binary system.

Our input can generally be represented as a vector or matrix, corresponding to each of our neurons (each blue dot in figure \ref{fig:neuralnetwork}) doing a matrix operation, as shown in this equation,

\begin{equation}
  \vec{y} = A(\omega \cdot \vec{x} + b)
\end{equation}

where $\vec{x}$ is the input, multiplied by weight $\omega$, which is a matrix and added by an offset $b$, then the nonlinear function $A$ acts on the result. When we train the machine, we are actually fitting this $\omega$. There are thousands of parameters in each $\omega$, and the whole network has hundreds of neurons, so the total number of parameters reaches the order of a million. To fit such a large number of parameters, a large amount of data input is required. The more data input, the more accurate the fit, the better the prediction ability of the machine. Usually we say the quality of data defines the upper limit of the accuracy, and the algorithm of the network always tries to improve the accuracy to reach the upper limit. 

Considering that we have a network and many data, how can we train the machine? Here we take supervised learning as an example, which is the most common training method at this stage. Supervised learning requires that the data provided has the correct answer. It is like teaching a child to distinguish a dog from a cat, you show him a cat picture, and he answers dog, then you correct him ``you are wrong, it should be a cat''. After many times of repeated training, the child masters the method of distinguishing between cats and dogs. Mathematically expressed, it is to find a set of parameters $\omega$ that makes the neural network output the same answer as the real answer, that is, the loss function is close to 0. Figure \ref{fig:gradientdescent} represents the loss corresponding to a certain $\omega$. The initial $\omega$ corresponds to a very high loss in the red region. For each input of data, or each training, we seek the gradient of the loss, our $\omega$ moves along the gradient by one small step in the inverse direction, with the step size controlled by $\eta$. After many training sessions, $\omega$ step by step eventually goes to a very small loss, the blue valley on the figure.

\begin{figure*}
\centering
\includegraphics[width=1\hsize,clip]{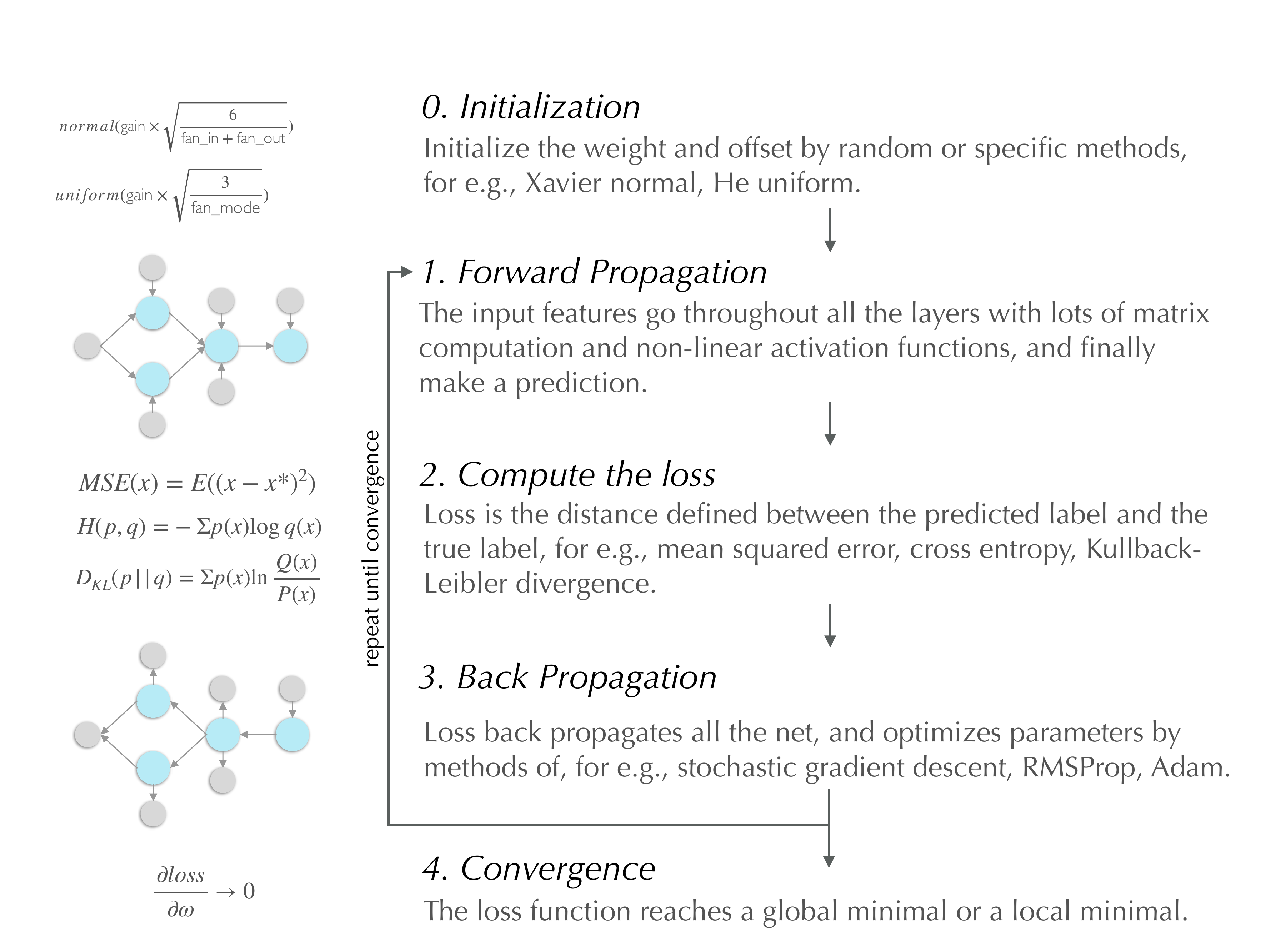}
\caption{Normally, a complete training process includes initiation, forward propagation, loss computation, backpropagation and the final convergence. The left side shows some examples of each step.}
\label{fig:fullprocedure}
\end{figure*}

\begin{figure*}
\centering
\includegraphics[width=1\hsize,clip]{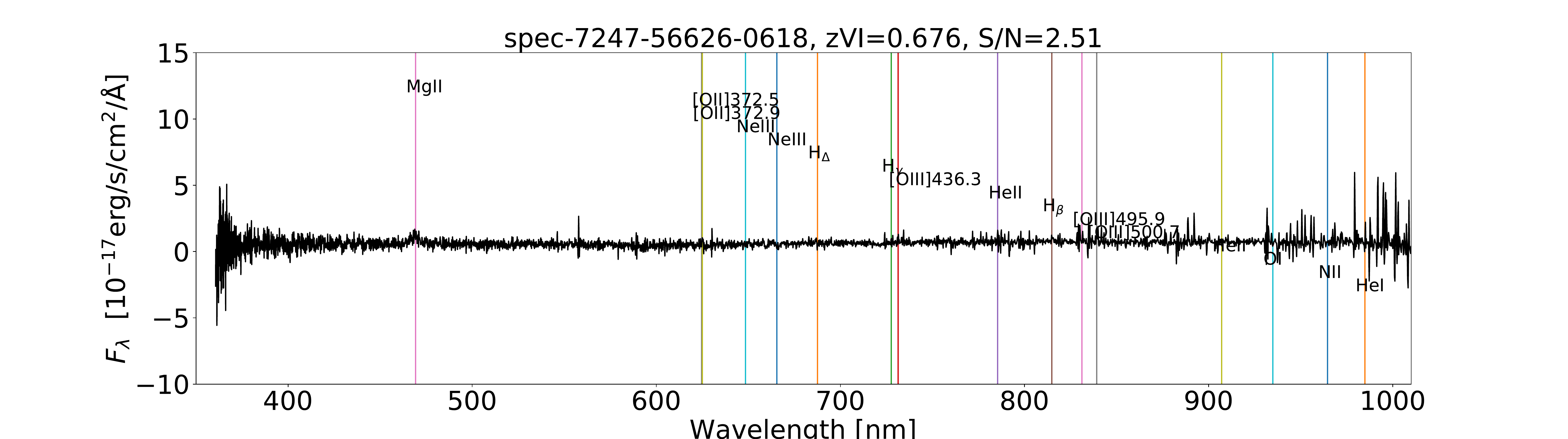}
\includegraphics[width=1\hsize,clip]{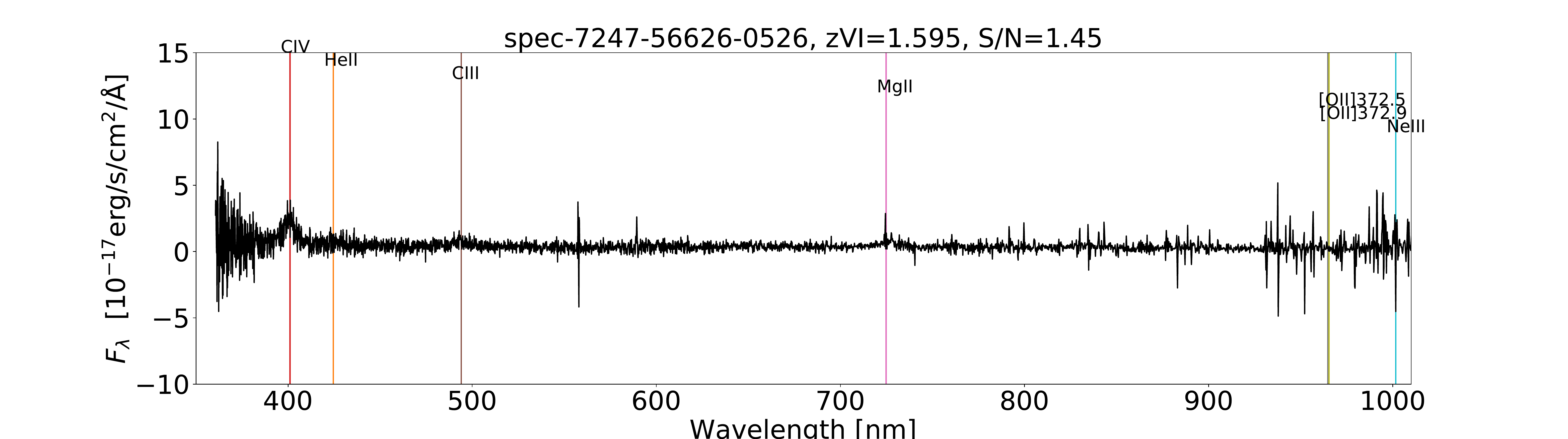}
\caption{The spectra of quasars. There are several emission/absorption lines present in the spectra, including Ly$\alpha$ (121.6 nm), $\rm CIV$ (154.9 nm), $\rm CIII$ (190.9 nm), $\rm MgII$ (279.6 nm), $\rm H\beta$ (486.2 nm) and $\rm H\alpha$ (656.3 nm) as well as a $\rm CIV$ line with a broad absorption feature. The spectral number, visually inspected redshift and the signal to ratio are labelled on each figure, respectively. This figure is reproduced from \citet{2022MNRAS.511.4490R}.}  
\label{fig:examples}
\end{figure*}

We go over the procedure, as shown in figure \ref{fig:fullprocedure}. First we design a neural network and initialize the parameters. Then we do the forward propagation to make a prediction. By comparing with the ground truth, we obtain the loss. Then the error is back propagated to the entire network to refine the parameters. We repeat this training process by feeding a lot of data until convergence is reached.

\section{Example}\label{sec:examples}

\begin{figure*}
\centering
\includegraphics[width=1\hsize,clip]{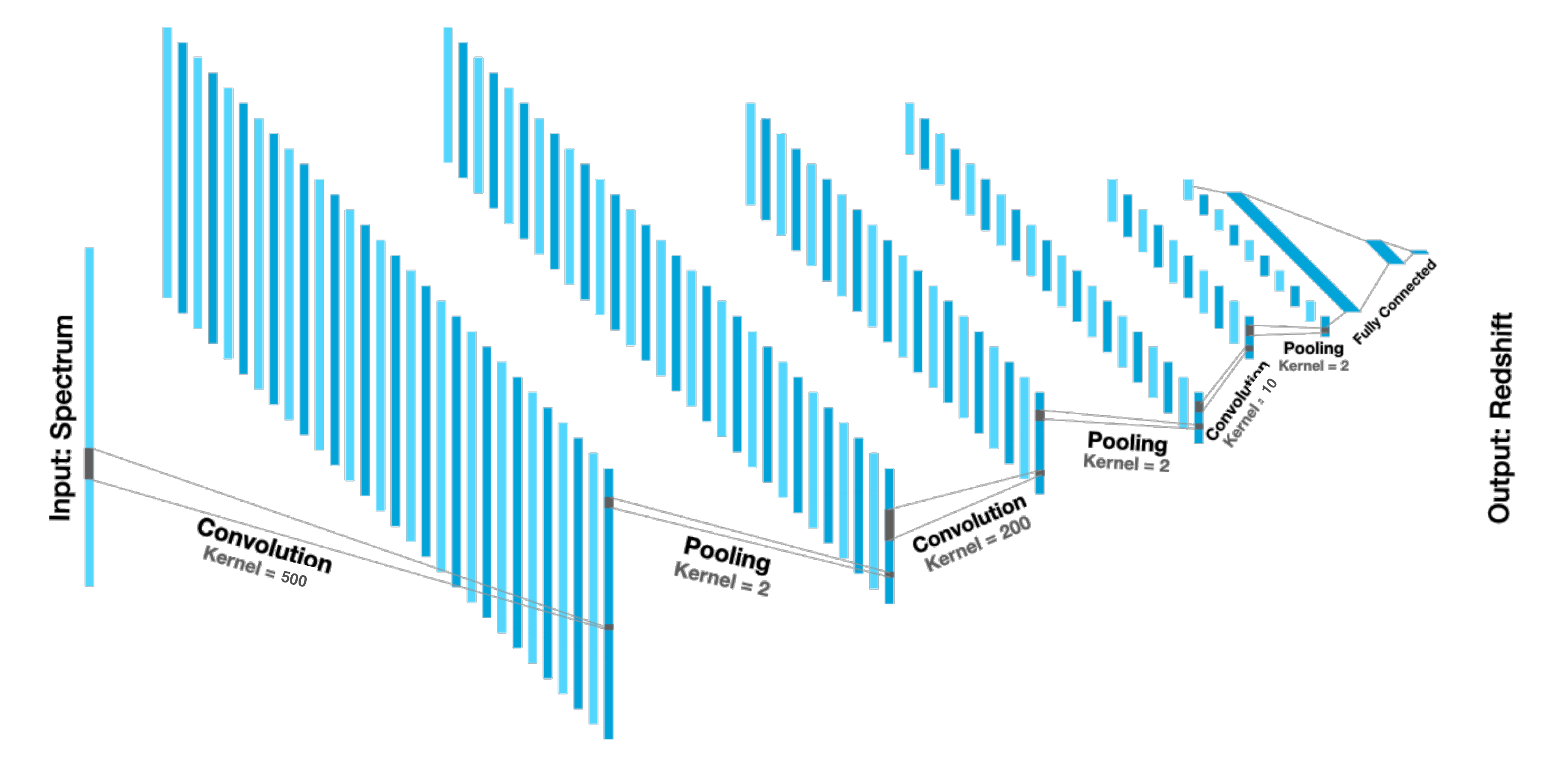}
\caption{Structure of one dimensional CNN. The quasar spectrum is input as a one-dimensional array, which goes through the convolutional layer of kernel size = 500, 200, 10 respectively to search for the global and local pattern. The fully connected layers output the redshift.}  
\label{fig:cnn}
\end{figure*}

Our example follows \citet{2022MNRAS.511.4490R}, that is to infer the spectroscopic redshift of quasars by deep learning. The relevant code and data can be found in \url{https://github.com/YWangScience/Isfahan-workshop-2021/} and \url{https://www.kaggle.com/datasets/ywangscience/sdss-iii-iv} respectively.

The quasar spectra are retrieved from Sloan Digital Sky Survey (SDSS) Data Release 16 and quasars only (DR16Q) \citep{2016AJ....152..205H,2020ApJS..250....8L}, which is one of the best samples to train the neural networks, with more than $700,000$ quasars, and includes $326,535$ quasars that are visually inspected. The spectra are standardized by fitting and then extrapolating to 4618 data points (pixels) uniformly distributed in $\log \lambda$, where $\lambda$ is the wavelength in the range of $360~\rm nm-1032.5~\rm nm$. Then the spectra are normalized using the Zero-Mean Normalization method \citep{jayalakshmi2011statistical} as features. The corresponding redshifts are exported as labels. We take $90\%$ of samples for training and $10\%$ for testing. Figure \ref{fig:examples} shows two spectra with the emission/absorption lines marked.
 
 In the field of machine learning, the most commonly used network architecture is the convolutional neural network (CNN), and many complex networks are built on top of the CNN. For our simple 1D data, a very deep network is rarely needed, and our experiments also demonstrate that building a CNN network with about $10$ layers is sufficient to obtain accurate redshift. We also tested that increasing the complexity of the network by a factor of 10 or more, or using a state-of-the-art network, yields an accuracy improvement of only $\sim 1\%$, so in this tutorial article we use the most basic CNN.

When designing CNN networks, we need to start from the actual physical problem. For the problem of deriving the redshift from the spectrum, we learn that the combination of these patterns can estimate the redshift:

\begin{enumerate}
  \item The global shift of spectrum;
  \item The shift of the emission and absorption lines at different redshifts;
  \item Some specific signals may appear at given redshifts. 
\end{enumerate}

Hence, in the convolutional part, we specially construct a large size filter of $500$ pixels covering $> 10\%$ of the data to capture the global shift of the spectrum, and in series a middle size $200$ pixels and a small size filter of $10$ pixels to capture the small and minor shift and those specific signals. The network architecture is shown in figure \ref{fig:cnn}.

The code of the whole process is written according to to figure \ref{fig:fullprocedure}, corresponding to:

\begin{enumerate}
  \setcounter{enumi}{-1}
  \item He uniform \citep{he2015delving} is adopted to initialize the network weights;
  \item forward propagation follows our net defined in figure \ref{fig:cnn};
  \item we adopt the mean squared error as the loss
  \begin{equation}
  	MSE(x) = E((x-x^*)^2)
  \end{equation}
  where $x$ is the prediction and $x^*$ presents the ground truth.
  \item Adam is selected to perform the backpropagation;
  \item we consider convergence when training cannot reduce the loss.
\end{enumerate}

After dozens of epochs of training, this simple network obtains an accuracy of $>97\%$ for the redshift of the testing samples.

\section{Extend the network by simple changes}\label{sec:extensions}

\begin{figure*}
\centering
\includegraphics[width=1\hsize,clip]{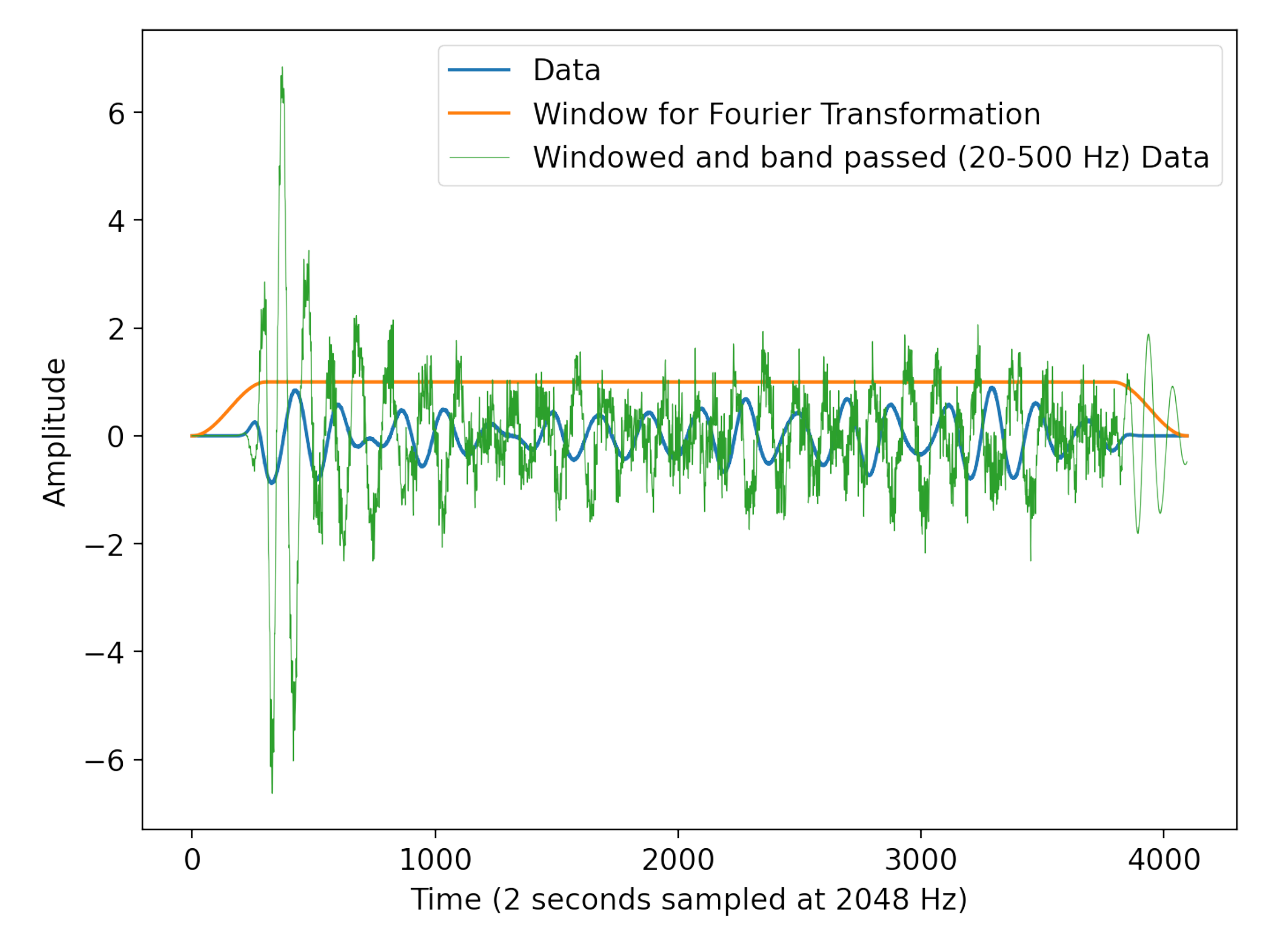}
\caption{Example of the gravitational wave data, the blue curve shows the original data, the orange curve is the window for Fourier transformation, and the green curve is the filtered signal that only contains $20-500$ Hz data.}  
\label{fig:gw}
\end{figure*}

Here by changing some lines of the net, we turn to train the net to classify the SDSS objects, and for simplicity of the tutorial, only involve two classes of quasar and galaxy.

This classification example uses the same input as the redshift example, a series of one-dimensional data. The outputs are different, The redshift example finally gives a redshift value, while here we output two values, corresponding to the probability of each class. In order to standardize the output, a LogSoftmax function
\begin{equation}
	p(x_{i}) = \log\left(\frac{\exp(x_i) }{ \sum_j \exp(x_j)} \right),
\end{equation}
 is applied at the last layer. Hence all values are between 0 and 1, and the sum of all equals one. For instance, the output ($0.2,0.8$) indicates that the source belongs to the second class. The loss function is also different, and the resulting $q(x_i)$ (predicted classification) will be adopted and together with the labels p(xi) (true ground classification) to compute the cross entropy as the loss 
 \begin{equation}
 	H(p,q) = \sum_i  p(x_i) \log q(x_i) 
 \end{equation}

With the above two simple changes to training the net following the same procedure of our first redshift example, this simple net is capable of making the classification of more than $80\%$ accuracy for quasar and galaxy.

With some simple modifications, this net can also be used for gravitational wave detection. Similar to the SDSS spectrum, the gravitational wave signal forms a one-dimensional data structure, that the time versus the strain, or the frequency versus the amplitude if changing from the time domain to the frequency domain, see figure \ref{fig:gw}. The output will be a value indicating the existence of gravitational wave. So we hardly need to modify the network structure,  just to modify the length of some parameters according to the length of the input data. Our test using the data provided from Kaggle competition (\url{https://www.kaggle.com/c/g2net-gravitational-wave-detection/}) shows such a simple net obtains an accuracy of more than $>80\%$, only about $\sim 5\%$ less accurate than the top one network. We also noticed in this competition that the top three networks did not use very complex networks but put much thought into pre-processing the data.

\section{Conclusion}\label{sec:conclusion}
Analyzing and processing data is the backbone of astronomy, and machine learning has arguably advanced the most. As a relatively new concept for astronomy, machine learning may be difficult for astronomers to start with. It occurred to us to illustrate through this article that, thanks to the easy-to-use programming frameworks developed by the industry, machine learning in astronomy does not require very difficult programming theories and capabilities. In fact, the main development of machine learning in the industry is in the processing of images and voices, which are actually recorded in a form no different from astronomical data, both in temporal and spatial data formats. These similarities allow us to apply many of the neural networks matured in the industry to astronomical data processing. In conclusion, machine learning as a tool is easy to use, but astronomers and astrophysicists need brilliant ideas to use this tool flexibly.

% ...


\bibliographystyle{aasjournal}
\bibliography{ml}

\end{document}